\documentclass[12pt,preprint]{aastex}

\slugcomment{Submitted: AJ, August 2011; Accepted September 2011|}

\shorttitle{Carnegie Hubble Program}
\shortauthors{Freedman et al.}

\begin{document}


\medskip
\medskip
\medskip

\title{\bf The Carnegie Hubble Program }
\medskip
\medskip
\medskip


\author{\bf Wendy L. Freedman, Barry F. Madore, Victoria Scowcroft, \\Andy Monson, S. E.  Persson, Mark Seibert}  
\affil{Observatories of the Carnegie Institution of Washington \\ 813
Santa Barbara St., Pasadena, CA ~91101}
\author{\bf Jane  R. Rigby}
\affil{Observational Cosmology Lab, NASA Goddard Space Flight Center, Greenbelt MD 20771}
\author{\bf Laura Sturch}
\affil{Department of Astronomy, Boston University, 725 Commonwealth Ave., Boston, MA, ~02215}
\author{\bf \&}
\author{\bf Peter Stetson}
\affil{Dominion Astrophysical Observatory \\ National Research Council of Canada \\
Herzberg Institute of Astrophysics, 5071 West Saanich Road \\ Victoria, BC V9E~~2E7}
\email{wendy@obs.carnegiescience.edu, barry@obs.carnegiescience.edu,
vs@obs.carnegiescience.edu, amonson@obs.carnegiescience.edu,  \\
persson@obs.carnegiescience.edu, mseibert@obs.carnegiescience.edu, \\
Jane.R.Rigby@nasa.gov, lsturch@bu.edu, Peter.Stetson@nrc-cnrc.gc.ca}


\medskip

\clearpage

\begin{abstract}
We present an overview of and preliminary results from an ongoing
comprehensive program that has a goal of determining the Hubble
constant to a systematic accuracy of $\pm$2\%.  As part of this
program, we are currently obtaining 3.6 $\mu$m data using the Infrared
Array Camera (IRAC) on {\it Spitzer}, and the program is designed to
include {\it JWST} in the future.  We demonstrate that the
mid-infrared period-luminosity relation for Cepheids at 3.6 $\mu$m is
the most accurate means of measuring Cepheid distances to date.  At
3.6 $\mu$m, it is possible to minimize the known remaining systematic
uncertainties in the Cepheid extragalactic distance scale.  We discuss
the advantages of 3.6 $\mu$m observations in minimizing systematic
effects in the Cepheid calibration of H$_\circ$ including the absolute
zero point, extinction corrections, and the effects of metallicity on
the colors and magnitudes of Cepheids.  We are undertaking three
independent tests of the sensitivity of the mid-IR Cepheid Leavitt Law
to metallicity, which when combined will allow a robust constraint on
the effect. Finally, we are providing a new mid-IR Tully-Fisher
relation for spiral galaxies.

\end{abstract}


\section{Introduction}
\subsection{The Need for Higher Accuracy in  H$_o$}

The determination of cosmological parameters has improved dramatically
in the past decade.  With measurements from WMAP, the HST Key Project,
Type Ia Supernovae and Baryon Acoustic Oscillations, to name just a
few examples, a new concordance cosmological model has emerged having
an expansion rate h = 0.72, density parameters $\Omega_m$ = 0.23, and
an equation of state parameter $w_{\circ}$ = -1.0, with an uncertainty
of $\pm$10$\%$ (Freedman \& Madore 2010).  The ``factor-of-two
controversy'' over the value of the Hubble constant has been resolved
(Freedman {\it et al.}  2001; Spergel {\it et al.}  2007; Riess {\it
  et al.}  2009) and 5-10\% or better precision has now been reached
for several fundamental cosmological parameters (Freedman \&
Madore 2010; Riess {\it et al.}  2011); Beutler et al. (2011), Komatsu
et al. (2011) and Mould (2011); a situation hard to imagine even just
a decade ago.


Equally impressive progress is expected given upcoming facilities such
as {\it LSST, EUCLID, GAIA} and the already-active mission
{\it Planck}. These large missions, combined with a suite of
small-to-medium sized projects planned for the interim, promise an
extraordinary opportunity to characterize and understand the processes
that govern the origin and evolution of the Universe.  However, strong
physical degeneracies exist amongst the cosmological parameters
derived from the angular power spectrum of cosmic microwave background
anisotropies (from {\it WMAP} and {\it Planck}, and other CMB
experiments). For example, there are well-known degeneracies between
the Hubble constant and other cosmological parameters, such as the
dark energy density $\Omega_\Lambda$ and $w_{\circ}$ (Hu \& Dodelson
2002).

An independent measurement of H$_o$, made to higher accuracy than we
have today, will continue to be a critical input for the next
generation of cosmology experiments.  Uncertainty in H$_o$, if left at
the 5\% level, will dominate the (coupled) uncertainties in the new
higher-precision experiments designed to measure cosmological
parameters. As emphasized by Hu (2005), the best complement to current
and future CMB measurements for a measure of the dark-energy equation
of state at a redshift of about z = 0.5 is a measurement of the Hubble
constant that is accurate at the few percent level. The results of
Beutler et al. (2011), Komatsu et al. (2011) and Mould (2011) are also
interesting in this regard.

\subsection{H$_o$ to 2\% Accuracy}

We are executing a program to recalibrate the extragalactic distance
scale and improve our knowledge of the Hubble constant by dealing
directly with all of the systematics currently known to be affecting
the Cepheid distance scale. We describe here our program using the
{\it Spitzer Space Telescope} (Werner{\it ~et al.} 2004) and Infrared
Array Camera (IRAC) (Fazio{\it ~et al.} 2004) as part of a {\it
  Spitzer} Exploration Project (PID 61000: Freedman).  The second
phase of our program (using more distant galaxies, deeper into the
Hubble flow) is designed to be carried out using the James Webb Space
Telescope ({\it JWST}).

The largest uncertainty in the HST Key Project determination of H$_o$
(Freedman{\it ~et al.} 2001) was the absolute zero point of the
Cepheid Period-Luminosity relation, which was directly tied to the
distance to the Large Magellanic Cloud (LMC).  Our first requirement
is to reduce the uncertainty in the absolute zero point by a factor of
two to three, using {\it Spitzer} to obtain complete lightcurve
coverage at mid-infrared (IRAC) wavelengths for a well-observed sample
of LMC and Galactic Cepheids. From there, we can use {\it Spitzer} to
extend these mid-infrared measurements to a much larger sample of HST
Key Project spiral galaxies in which Cepheids have already been
discovered.  This single step alone immediately drops the systematic
error in the Cepheid distance scale zero point to the (3\%) level.  As
discussed in Freedman {\it et al.} (2012) and briefly in \S
\ref{sec:H0}, a preliminary calibration based on our {\it Spitzer}
data already exceeds the Key Project in accuracy by over a factor of
three. As a cross check, we are also measuring the distance to the
maser galaxy, NGC 4258, which is itself an independent calibrator of
the distance scale zero point, in addition to the LMC and the Milky
Way (Humphreys {\it et al.} 2008; Mager {\it et al.} 2008; Riess {\it
  et al.} 2011). Our recent results, discussed below, demonstrate that
we can reach this required level of accuracy. We will then be in a
position to measure distances into the Hubble flow, this time by using
IRAC to both produce a new mid-IR Tully-Fisher relation for spiral
galaxies.  Future measurements using {\it JWST} will then
significantly increase the number of fundamental Cepheid calibrators
and bring the uncertainties down to $\pm$2\% We discuss each of these
steps in turn, and give an overview of some preliminary results based
on our current analysis of {\it Spitzer} data.

Two relatively recent developments have dramatically changed the
landscape regarding a recalibration of the extragalactic distance
scale.  First, Benedict {\it ~et al.} (2007) used the Fine Guidance
Sensors (FGS) on HST to provide the first high-precision, geometric
parallaxes to 10 nearby Galactic Cepheids having periods ranging from
4 to 35 days.  Second, Freedman {\it ~et al.} (2008) demonstrated (using
{\it Spitzer}/SAGE legacy data for the LMC from Meixner {\it et al.}
2006) the small dispersion in the mid-infrared Period-Luminosity (PL)
relations (hereafter referred to as the Leavitt Law) at
mid-infrared wavelengths, even for single (not phase-averaged)
observations of Cepheids.

\section{Why the Mid-Infrared?}
\label{sec:midIR}

Mid-infrared observations of Cepheids offer a host of advantages over
shorter wavelength data. Most important is the reduced sensitivity of
long-wavelength data to line-of-sight interstellar extinction (both
Galactic and extragalactic). The interstellar extinction law at
mid-infrared wavelengths has now been measured by a number of authors
(Rieke \& Lebofsky 1985; Indebetouw{\it ~et al.} 2005; Flaherty{\it
  ~et al.} 2007; Roman-Zuniga{\it ~et al.} 2007; Nishiyama{\it ~et
  al.}  2009).  They find that the shape of the extinction curve
varies somewhat between different sightlines, where the observed range
of $A_{\lambda}/A_V$ is 0.058--0.071 at 3.6 $\mu$m, and 0.023--0.060
at 4.5 $\mu$m.  However, the most important point is that the
extinction measured in magnitudes in the mid-IR, as compared to
optical $V$-band data, for example, is reduced by factors of 14-17 at
3.6 $\mu$m, and 16-43 at 4.5 $\mu$m.

In practice then, for $A_V \sim$ 0.20~mag, with an uncertainty in the
reddening of 10\% ($\epsilon$[$E(B-V)$ =  $\pm$0.02) mag (giving
$\epsilon$[A$_V$] of $\pm$0.06 mag), at mid-IR wavelengths this
reduces to $A_{3.6\mu m}$ = 0.01$\pm$ 0.004~mag; that is, 
 the correction to a V-band distance transforms from 10\% to a
correction of only 0.5\% in distance at 3.6 $\mu$m,  and it drops the
uncertainty on the distance from 3\% in the visual down to a
statistically insignificant level of only $\pm$0.2\% at 3.6~$\mu$m.

In addition, because stellar surface brightness in the mid-infrared is
so insensitive to temperature (being on the Rayleigh-Jeans portion of
the spectral energy distribution), the observed, cyclical variation in
a Cepheid's luminosity at 3.6 $\mu$m is almost completely dominated by
the comparatively small radial (i.e., surface area) variations. The
slope of the Period-Luminosity relation in the mid-IR then becomes virtually
equivalent to the Period-Area relation, when the bolometric
correction becomes insensitive to temperature (which occurs longward
of about one micron for G and K spectral types, typical of Cepheids.)
The areal variations typically amount to only around 0.3-0.4~mag in
total amplitude. (Those over 100 days can have amplitudes of 0.6-0.8
mag). This is to be compared with observed B-band amplitudes that can
exceed 2~mag or I-band amplitudes that can reach 1~mag.  Even just a
factor of three decrease in amplitude provides for almost a factor of
10 decrease in the number of randomly-phased observations needed to
reach the same error on the mean magnitude for a given variable star.

There are yet further advantages to the mid-infrared. The effects of
line blanketing, the process by which energy is removed from the
optical and UV and is subsequently thermalized and redistributed to
the optical/near-infrared (back-warming), are expected to be much
smaller at mid-infrared wavelengths\footnote{However, at cooler
  temperatures the formation of molecules and the appearance of
  molecular bands, such as CO in the 4.5$\mu$m region do run contrary
  to this overall expectation.}. Most importantly, because the
reddening is so low in comparison to the optical, 3.6 $\mu$m provides
an opportunity for a very clean and direct test for any metallicity
effects.


For the HST Key Project determination of the extragalactic distance
scale, a list of identified systematic uncertainties set its finally
quoted accuracy at $\pm$10\%. The dominant systematics were: (a) the
zero point of the Leavitt Law; (b) the differential metallicity
corrections to the PL zero point; (c) reddening corrections; (d)
calibration/instrumental uncertainties; (e) crowding effects and
finally HST PSF uncertainties and evolution with time/position. In
this paper, we discuss the improvements that are coming from {\it
  Spitzer}, and that will eventually come from JWST.

\subsection{Model Spectra of Cepheids from 4-6 $\mu$m}

We show in Figures 1a and b Kurucz models for solar-metallicity
supergiant stars, with effective temperatures and gravities of
4000K and log(g)=0, consistent with those of Cepheids. Note that
the IRAC band 1 at 3.6 $\mu$m is devoid of any molecular bands;
however, IRAC band 2 at 4.5 $\mu$m overlaps with the broad CO molecular
bands in the approximate wavelength range of 4 to 6 $\mu$m. There are two immediate
conclusions from these plots. First, Figure 1a illustrates the smooth
continuum over the 3.6 $\mu$m band, and as expected, its suitability
for distance determinations. Figure 1b illustrates that the 4.5 $\mu$m
band may not be a suitable distance indicator for Cepheids. We are
currently exploring the use of the 4.5 $\mu$m band as a metallicity
indicator for Cepheids (Scowcroft {\it et al.}, in
preparation). 

\begin{figure*}
    \centering
    \includegraphics[width=5.5cm, angle=-90]{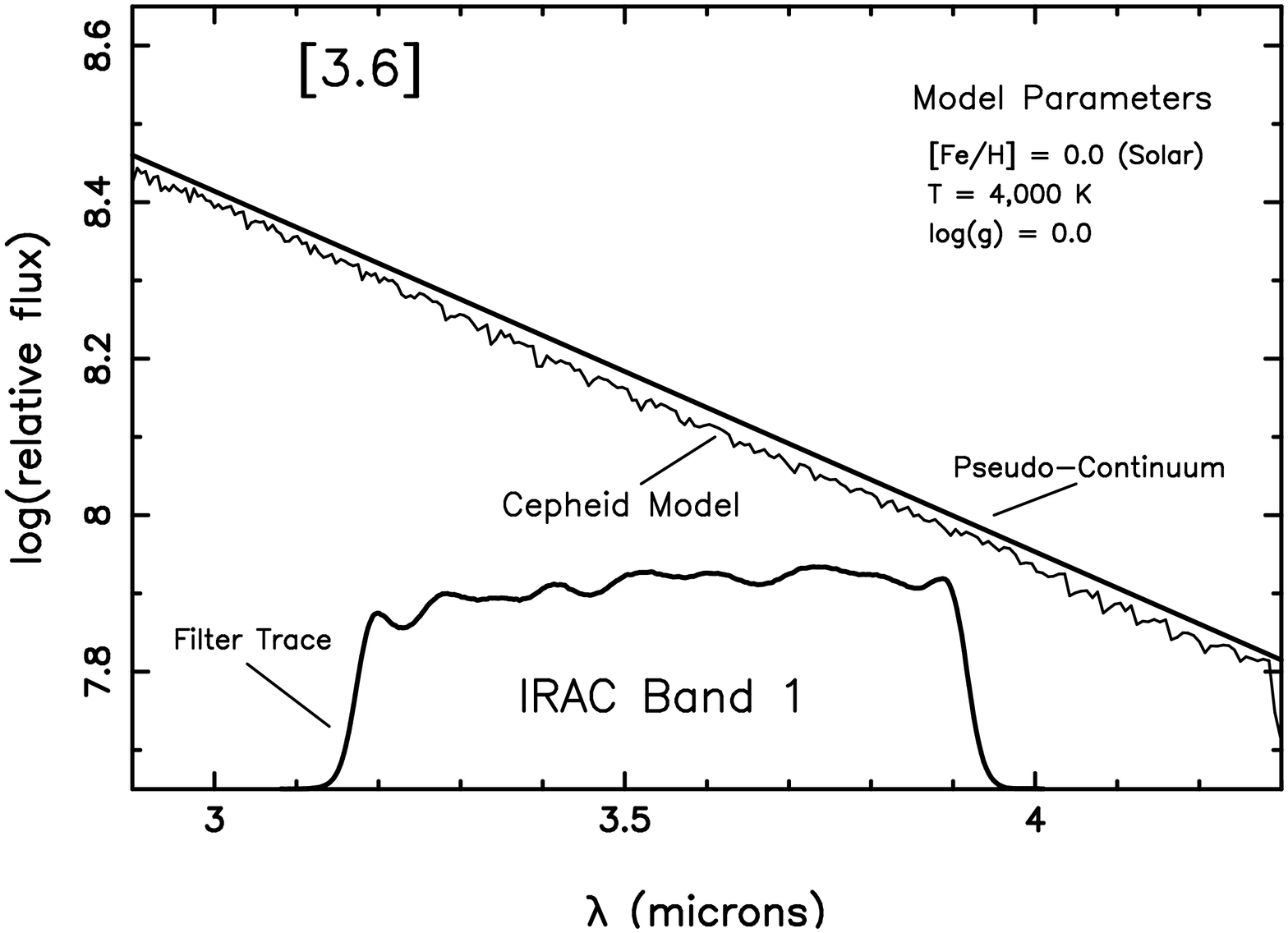} 
    \includegraphics[width=5.5cm, angle=-90]{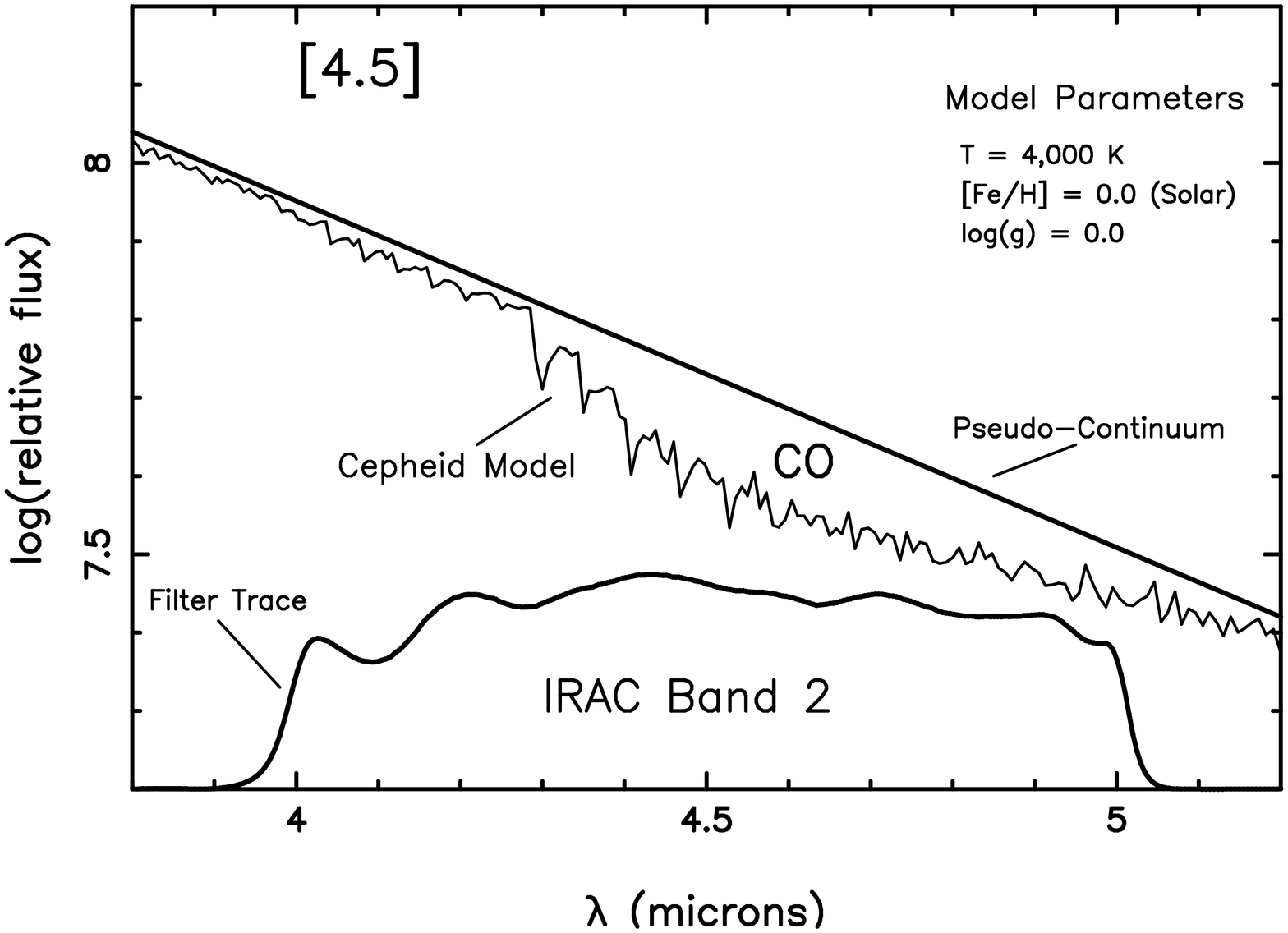} 
\caption{\small Synthetic spectra for supergiants of solar
  metallicity, spanning the wavelength ranges for the 3.6 and 4.5 
  $\mu$m bands respectively.  These plots were generated by AM using
  the code and line lists from
  http://wwwuser.oat.ts.astro.it/atmos/Download.html (Kurucz 1993;
  Sbordone {\it et al.} 2004, 2005). The spectra are shown at an
  effective spectral resolution of R=600.  }
\end{figure*}

These results are consistent with those obtained by Marengo {\it et
  al.} (2010) based on time dependent hydrodynamic models of
Cepheids. Their Figure 6, based on a model for the 10-day Cepheid,
$\zeta$ Gem, shows a broad CO feature at 4.5 (as well as 5.8) $\mu$m,
but no features at 3.6 $\mu$m. This behavior is also consistent with
the observed [3.6]-[4.5] $\mu$m colors of Cepheids (e.g., Marengo {\it
  et al.}, Monson {\it et al.} 2012; Scowcroft {\it et al.} 2012a);
see also \S \ref{sec:CO}.  Finally, we see the effects of the CO bands
at 4.5 $\mu$m in the behavior of the Leavitt Law slopes as a function
of wavelength. The slopes approach an asymptotic value of -3.45;
however, the 4.5 $\mu$m slope is clearly discrepant (e.g., Freedman
{\it et al.}  2008; Marengo {\it et al.} 2010; Scowcroft {\it et al.}
2012a). This body of evidence suggests that the 4.5 $\mu$m band may
need to be avoided for distance determinations.

\section{The Carnegie Hubble Program (CHP)}

We give here a brief overview of the components of the CHP observing
program.  The galaxies for which we have used {\it Spitzer} to obtain
observations of Cepheids are given in Table 1, which lists all of the
targets by program, the number of observations per Cepheid, how the
observations were spaced (phased or random), and the filters used. We
also list the numbers of Tully-Fisher calibrators and Tully-Fisher and
SNe~Ia target galaxies.  All of our observations have been made using
post-cryogenic or ``Warm {\it Spitzer}''.

Originally we obtained 3.6 and 4.5 $\mu$m measurements of 37 Galactic
Cepheids (Monson {\it et al.} 2012). Each Cepheid was observed 24
times over the course of its cycle, and the observations were
scheduled to give uniform sampling of the light curves. All of these
Cepheids are close enough to be future GAIA satellite targets, which will
provide accurate parallaxes for a larger sample of Milky Way Cepheids.
We have been awarded further {\it Spitzer} time to observe a larger sample
of Galactic Cepheids, in anticipation of the launch of GAIA, which
would bring the number of Milky Way targets in line with our LMC and
SMC samples.  Ultimately this larger sample of Cepheids will provide a
robust zero point for the calibration of H$_{\circ}$. We have also
obtained twenty-four 3.6 and 4.5 $\mu$m measurements of 85
well-observed Cepheids in the LMC (Scowcroft {\it et al.} 2012a).  As
an adjunct program, we have also obtained 12 epochs of 3.6 and 4.5
$\mu$m measurements for a sample of 100 SMC Cepheids, useful for
calibration of the metallicity effects for Cepheids.

In Figure 2 we show 3.6 and 4.5 $\mu m$ example light curves for
two Galactic Cepheids from Monson {\it et al.} (2012), two LMC
Cepheids from Scowcroft {\it et al.} (2012a), and two SMC Cepheids
from Scowcroft {\it et al.} (2012b).  These Cepheids have a range of
  periods from 5  to 66  days. The very small scatter in these light curves
  indicates the high quality of the {\it Spitzer} photometry.

\begin{figure}
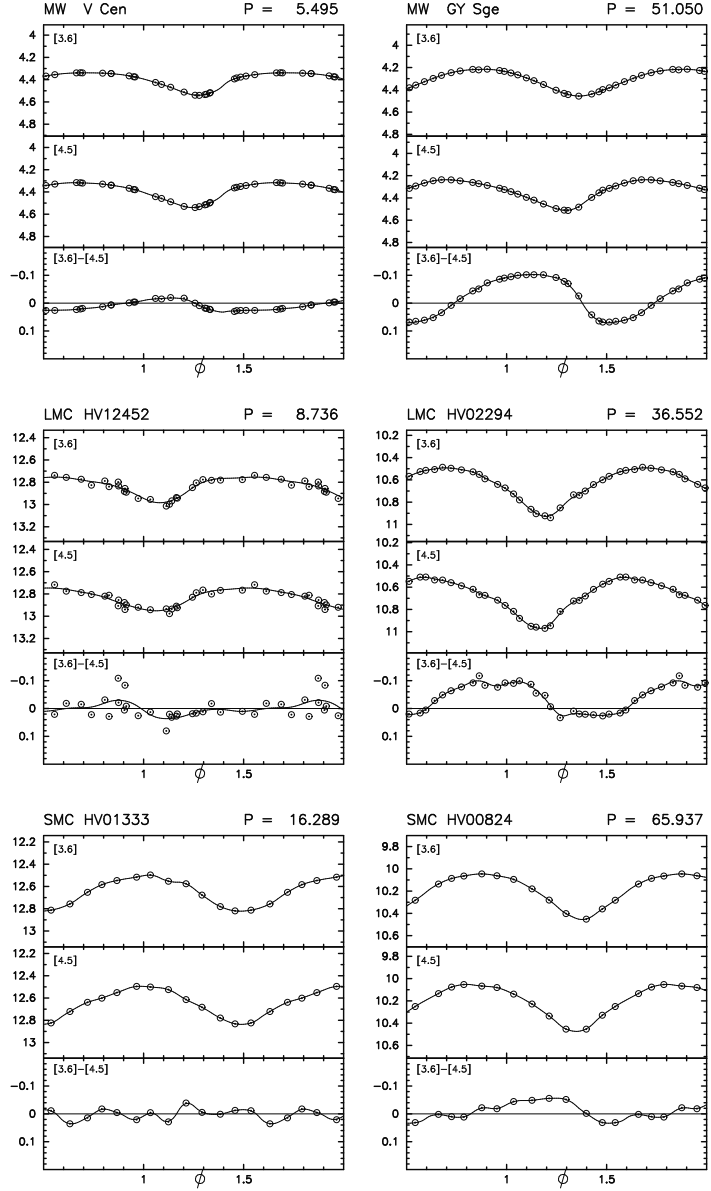
 
 \begin{center}$ 
 \begin{array}{cc} 
\includegraphics[width=50mm,angle=270]{f2a.eps} & 
\includegraphics[width=50mm,angle=270]{f2b.eps} \\
\includegraphics[width=50mm,angle=270]{f2c.eps} & 
\includegraphics[width=50mm,angle=270]{f2d.eps} \\
\includegraphics[width=50mm,angle=270]{f2e.eps} &
\includegraphics[width=50mm,angle=270]{f2f.eps} \\ 
\end{array}$ 
\end{center} 
\label{fig:lightcurves}\caption{Light and color curves for a selection of Cepheids in the Milky Way
  (top row), LMC (middle row) and SMC (bottom row). Cepheids with
  periods longer than 12 days are observed at regularly spaced
  intervals of $P/24$ (or $P/12$ for the SMC). Short period
  ($P<12$ days) Cepheids such as V Cen are observed every $12 \pm 4$
  days. These plots illustrate the high-quality,  well-sampled light
  curves that can be obtained  over a large range of periods  utilizing the outstanding
  scheduling efficiency of \textit{Spitzer}.}
\end{figure}

Moving out in distance, we have obtained 12 epochs of 3.6 $\mu$m data
for several galaxies within the Local Group and beyond containing
known Cepheids (see Table 1).  Critical for an independent zero-point
calibration of the Leavitt Law, we have observed Cepheids in the maser
galaxy NGC 4258 (at 7.2~Mpc).  Finally, we have obtained 3.6 $\mu$m
photometry for several hundred galaxies located in clusters with
measured Tully-Fisher distances, which can then be calibrated with
Cepheids (\S \ref{sec:TF}).  Over 50 galaxies with SNe~Ia distances
measured by Contreras {\it et al.} (2009) have also been observed as
part of this program, which will allow an independent determination of
H$_o$ with this calibration well into the far-field Hubble flow. We
now describe the individual aspects of the CHP in more detail.

\subsection{The Galactic Cepheid Calibration}

To date, as part of the Carnegie Hubble Program to set the absolute
zero-point and slope of the Leavitt Law relation in the mid-IR, we
have obtained high-precision, uniformly-sampled, 3.6 and 4.5 $\mu$m
light curves for the Galactic Cepheid parallax sample (Monson ~et
al. 2012). All of the light curves have comparable quality to what is
shown in Figure 2.  We also have equivalent data for an
order-of-magnitude larger sample of LMC Cepheids (Scowcroft ~{\it et
  al.} 2012a), discussed further below.  The former calibrates the
absolute zero point; the latter determines the slope. In addition to
the 10 Milky Way Cepheids with HST trigonometric parallaxes from the
FGS (Benedict{\it ~et al.} 2007), we have observed 27 other Cepheids
within 4~kpc of the Sun (Monson {\it et al.} 2012) that are close
enough for astrometric parallaxes for GAIA.  The currently available
sample of $HST/FGS$ Galactic Cepheids yield a zero-point calibration
of the Cepheid PL relation good to $\pm$2\%.  An independent check of
this calibration and a 50\% statistical improvement will come with the
{\it Spitzer} observations of nearby Cepheids once GAIA has obtained
geometric parallaxes for them.  Seventeen of the Cepheids in our
sample are also known to be members of Galactic open clusters or
associations (Turner 2010), for which GAIA again will provide a more
accurate calibration.

\subsection{The Distance to the LMC}

Distance measurements to the Large Magellanic Cloud have played a
critical role in the calibration of the extragalactic distance scale
({\it e.g.,} Freedman {\it et al.} 2001, Riess {\it et al.} 2005,
Sandage {\it et al.} 2006).  Because of its proximity, a number of
different methods have been used to estimate the distance to the
LMC.\footnote{See the extensive compilation of 275 distance estimates
  to the LMC currently available from the NASA/IPAC Extragalactic
  Database~(NED):
  http://nedwww.ipac.caltech.edu/cgi-bin/nDistance?name=lmc} The range
in quoted distances is almost certainly dominated by systematic
errors. As tabulated in Gibson (2000) and in Freedman {\it ~et al.}
(2001), most of the values for the LMC distance modulus have tended to
fall between 18.1 and 18.7~mag (i.e., 42 to 55~kpc); more recent
values have tended to cluster around a distance modulus of 18.5~mag
(see Alves 2004, and Schaefer 2007) which is the value adopted by the
Key Project.

In the LMC there are 92 Cepheids for which there are published optical
(BVRI) and near-infrared (JHK) light curves and time-averaged
photometry (Madore \& Freedman 1991; Persson{\it ~et al.} 2004). These
stars were chosen to be unconfused in the K-band, distributed across
the face of the LMC, and having periods ranging from 3 to 100
days. Approximately two dozen near-IR phase points were obtained at
each (JHK) wavelength for each star. We have now obtained
time-averaged IRAC photometry for 85 of these LMC Cepheids (selected
to have periods in excess of 6 days), consisting of two dozen
observations at both 3.6 and 4.5 $\mu$m (Scowcroft {\it ~et al.}
2012a). These observations were scheduled evenly based on the known
period for each Cepheid so that they would be uniformly spaced when
phase-folded into a light curve, the success of which can be seen in
Figure ~2.  

Long-period Cepheids are intrinsically the brightest and are therefore
the first (and sometimes the only) Cepheids detected in the most
distant galaxies. The extragalactic distance scale therefore rests
heavily on the long-period Cepheid calibration. However, the Galactic
HST parallax sample contains only one long-period Cepheid, $l$~Car at
P = 35.5~days.  The final set of LMC Cepheids observed for the CHP is
almost an order of magnitude larger in size than the currently
available total set of Galactic calibrators, and is therefore being
used to define the slope and width of the long-period (P $\ge$ 10
days) end of the Leavitt Law.

In Figure ~3, we show the Leavitt Law at 3.6 $\mu$m for 82 Cepheids in
the LMC as observed by Scowcroft {\it et al.} (2012a), with 6 $<$ P
$<$ 60 days.  Cepheids with log P $>$ 1.8 are shown, but have been
excluded from the fits. The dispersion in the 3.6 $\mu$m relation
amounts to only $\pm$0.106 mag or $\pm$5\% in distance for a single
Cepheid. For comparison, we also show the V-band data from Madore \&
Freedman (1991) for LMC Cepheids. The dispersion in this case is more
than a factor of two greater, amounting to $\pm$0.252 mag. Comparison
with the dispersion seen the near-infrared is quite favorable, with
the H and K$_s$-band dispersions being $\pm$0.116~mag
and$\pm$0.108~mag, respectively (Persson et al. 2004).

\begin{figure*}
    \centering
    \includegraphics[width=14.0cm, angle=-0]{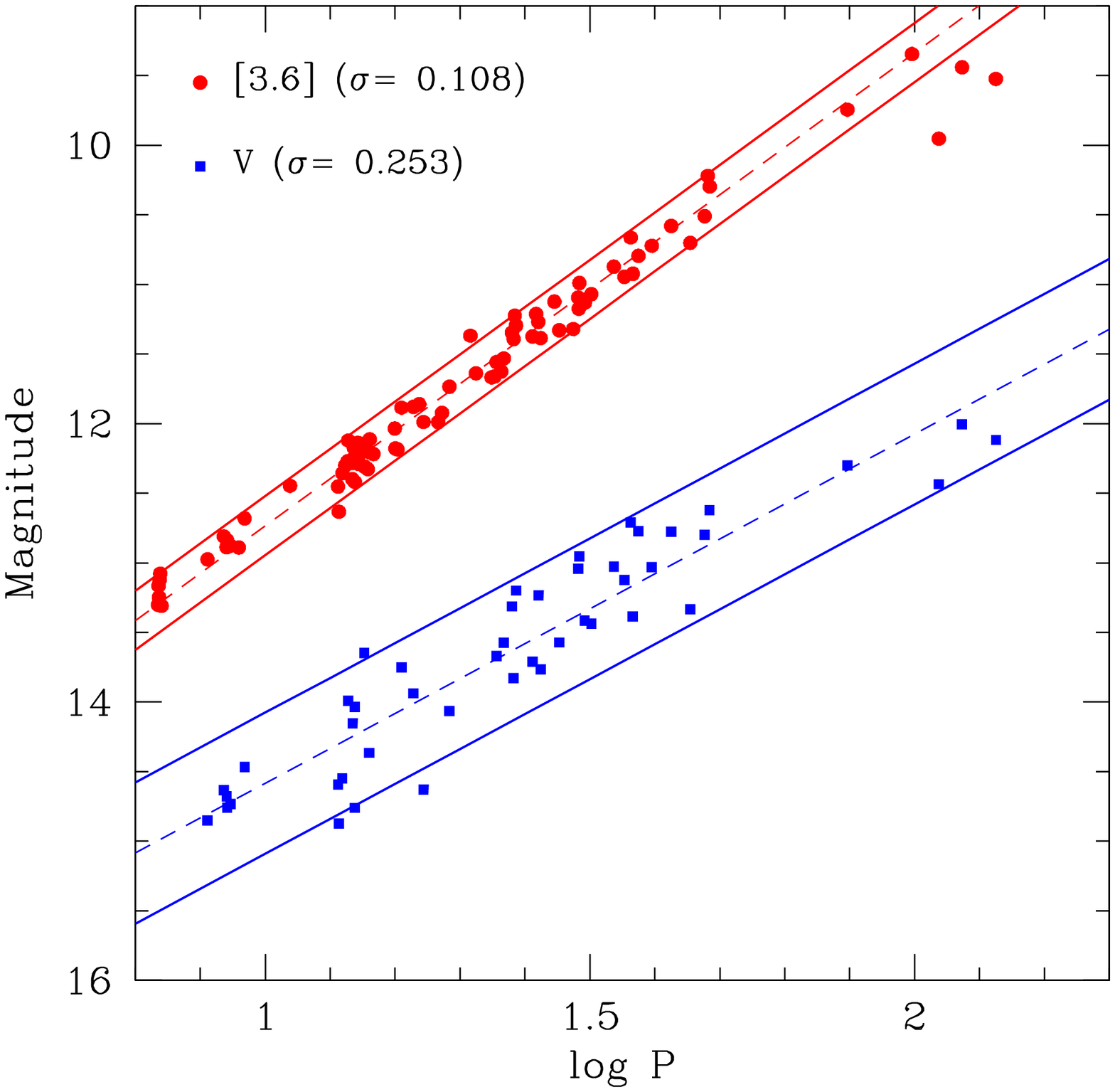}
\caption{ Phase-averaged 3.6 $\mu$m (red circles) and V-band (blue
  squares) Leavitt Law relations for the LMC. The 3.6 $\mu$m data are
  from Scowcroft {\it et al.} (2012a); the V-band from Madore \&
  Freedman (1991).  Note the small dispersion of $\pm$0.108~mag at 3.6
  $\mu$m, which is more than a factor of two less than for the
  V-band. The dashed lines represent weighted least squares fits to
  the PL relations for Cepheids in the period range 6 to 60 days. The
  solid lines denote 2-$\sigma$ ridge lines. }
\end{figure*}

These data demonstrate the power of {\it Spitzer} to provide a
quantitatively significant improvement in the calibration of the
Hubble constant based upon a Cepheid distance scale. As discussed
above, at 3.6 $\mu$m compared to optical wavelengths, the effects of
reddening are significantly lower, the expected effects of metallicity
are lower (see \S \ref{sec:metallicity}), and the dispersion in the
Leavitt Law is about $\pm$0.1 mag, giving distances good to $\pm$5\%
for a single Cepheid. Accordingly, for a sample of 82 Cepheids, the
distance to the LMC can be determined to a level of statistical
precision better than $\pm$1\%.  The systematic accuracy is $\pm$2\%;
this lower limit being set by small numbers of Cepheids in the
Galactic calibration. With the addition of parallaxes from GAIA, we
can expect the number of Galactic calibrators to more than double,
thereby setting a final systematic uncertainty of 1.5\% in distance.

\subsection{Cepheid Distances to  Local Group Galaxies}

We are currently obtaining high-precision, mid-IR Cepheid distances to
all Local Group galaxies known to contain Classical Cepheids.  Several
of these are calibrators for the Tully-Fisher relation, while others
also can act as a basis for testing for metallicity effects on the
Cepheid PL relation (see below).  The Local Group targets are listed in Table 1.
All of these galaxies and their Cepheids have been observed at optical
wavelengths and many of them have already been followed up in one or
more near-infrared bands ({\it e.g.}, Gieren {\it et al.} 2008 and
references therein for the Aruacaria Project on near-infrared
observations of Cepheids in southern hemisphere galaxies).

We have analyzed publicly available archival data on two Local Group
galaxies, NGC~6822 (at 500~kpc; Madore{\it ~et al.} 2009) and IC~1613
(at 700~kpc; Freedman{\it ~et al.} 2009). In each of these galaxies we
were able to recover and measure the brightest Cepheids that were
uncrowded. Analysis of all 12 epochs of data for IC 1613 is currently
underway (Scowcroft {\it et al.} 2012b). The NGC~6822 PL
relations in all four IRAC bands are shown in the left panel of Figure
4.  These archival exposures were not optimized for measuring faint
stars, but they clearly show that Cepheids can be seen out to the edge
of the Local Group, and beyond. The exposure time per pixel for
NGC~6822 was only 240~sec, and yet Cepheids with periods from 10 to
100 days were easily measured.  The right panel of Figure 4
demonstrates the ability of IRAC observations to cut through the
intervening extinction. In this instance, the apparent B modulus is
affected by  1.1~mag of extinction; while the uncorrected
3.6 $\mu$m apparent modulus is within 0.08~mag of the  true
modulus.

\begin{figure*}
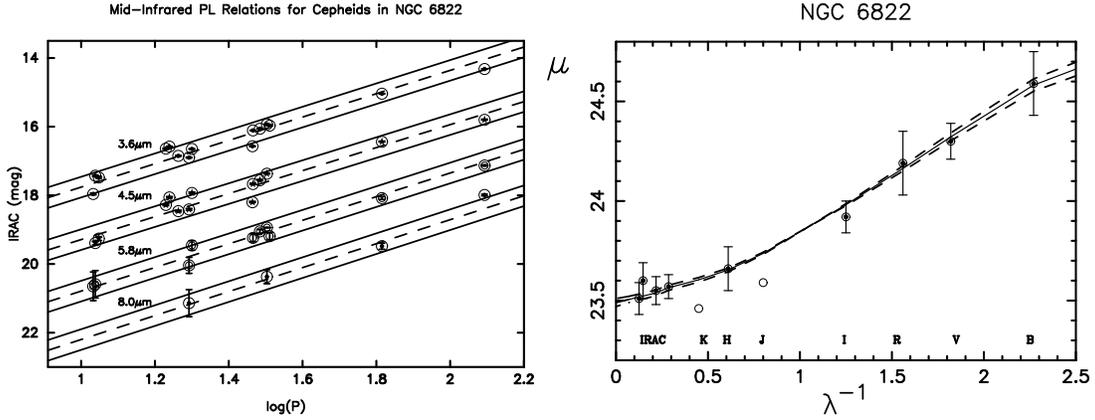

    \centering
    \includegraphics[width=5.5cm, angle=-90]{f4a.ps}
    \includegraphics[width=5.5cm, angle=-90]{f4b.ps}
\caption{\small Left Panel: Random-phase Leavitt Law relations for
  Cepheids in NGC~6822 observed as part of the SINGS Legacy program  in
  all four IRAC bands. Right Panel: Multiwavelength fit of a Galactic
  extinction curve to apparent Cepheid distance moduli to NGC
  6822. The data are from Madore {\it et al.}  (2009). The shape of
  the curve is fixed; its amplitude is set by the line-of-sight
  extinction; and the intercept gives the true distance modulus. The
  two near-infrared data points fall off of the main fit for currently
  unknown reasons. The mid-IR apparent distance moduli appear close to
  the true modulus even before any extinction correction is applied. }
\end{figure*}

\subsection{Beyond the Local Group}

As summarized in Table 1, we have obtained 3.6 $\mu$m imaging data for
a number of nearby galaxies with known Cepheids beyond the Local
Group.  The primary challenge for these more distant galaxies will be
the effects of crowding, or overlapping of images.  We are exploring a
number of methods to mitigate the effects of crowding on the
photometry.  These galaxies will serve the dual roles of calibrators
of secondary distance methods and probes of the expansion rate inside
8~Mpc or so. Beyond this distance, the higher resolution and higher
sensitivity of JWST will be required for accurate measurements of
Cepheid distances at 3.6 $\mu$m.  The goal of attaining a measurement
of H$_\circ$ to a (statistical and systematic) level of $\pm$2\% will
require direct Cepheid distances to the larger sample of more distant
galaxies. Most of the galaxies with Cepheid distances measured as part
of the Key Project, as well as the increasing samples of Type~Ia
supernovae ({\it e.g.,} Riess {\it et al.} 2011) will require JWST
observations.

\subsection{The Mid-IR Tully-Fisher Relation and the Far-Field Hubble Flow}
\label{sec:TF}

There are 10-20 galaxies in each of the clusters calibrated as part of
the HST Key Project sample, based on the survey of Giovanelli {\it
  al.} (1997), that can be used to determine $H_{\circ}$ using the
{\it Spitzer}-calibrated mid-IR Tully-Fisher relation. The galaxies in
these clusters probe 9,000~km/sec (or 120~Mpc) into the Hubble
flow. We have further supplemented this cluster sample with field
objects selected from the Flat Galaxy Catalog (Karachentsev,
Karachentseva \& Paranovsky 1993) for which there are published
HI-line profiles. This sample extends the TF reach to around
18,000~km/sec (or 240~Mpc). The two samples will allow us to average
over any residual perturbations to the Hubble flow even at high
redshift, and guard against possible environmental (field versus
cluster) effects.  We have also obtained 3.6 $\mu$m data for nine
galaxies which may ultimately have water maser distances independently
determined for them (Braatz {\it et al.} 2008). Finally, we are
observing a sample of disk galaxies drawn from The Carnegie (Low-z)
Supernova Program (Contreras{\it ~et al.} 2009), that have had Type Ia
supernova events measured in them, providing us with a direct
cross-calibration and tie-in of Tully-Fisher with the SN distance
scale. Analysis of these data is currently underway (Seibert {\it et
  al.} 2012).  A more direct tie-in between galaxies beyond the
resolution/confusion limit of {\it Spitzer}, having both Cepheids and
Type Ia SN events, must await for the combined sensitivity and
resolution of {\it JWST.}

In addition to Cepheids, {\it Spitzer} also offers advantages for the
Tully-Fisher relation in the mid-IR where once again the effects of
extinction are minimized. In addition, the contribution of old stars
as tracers of mass can be maximized. Of the 23 nearby galaxies that
can be used to calibrate the Tully-Fisher relation, having distances
determined by HST, 17 of these have new 3.6 $\mu$m AB magnitudes
(Seibert et al. 2012).  In Figure 5 we show the B, V, I and 3.6 $\mu$
sample of 17 calibrating galaxies for which there are data for all
four wavelengths. The B, V, and I-band magnitudes have been corrected
for inclination-induced extinction effects and their 20\% line widths
have been corrected to edge-on (Sakai{\it ~et al.}  2000); no
extinction correction has been applied to the 3.6 $\mu$m data. The
1-$\sigma$ scatter in these relations is $\pm$0.43, 0.37, 0.32 and
0.31 ~mag for the B, V, I and 3.6 $\mu$m data, respectively; the outer
lines follow the mean regressions at $\pm$2-sigma.  Each of these
galaxies entered the calibration with its own independently determined
Cepheid-calibrated distance from Freedman{\it ~et al.}  (2001). In
Figure 6, we show example Tully-Fisher 3.6 $\mu$m relations for 4 out
of 24 clusters of galaxies for which we have 3.6$\mu$m data. These
data will provide an independent estimate of the value of H$_\circ$
(Seibert {\it et al.} 2012).

\begin{figure*}
    \centering
    \includegraphics[width=10cm, angle=0]{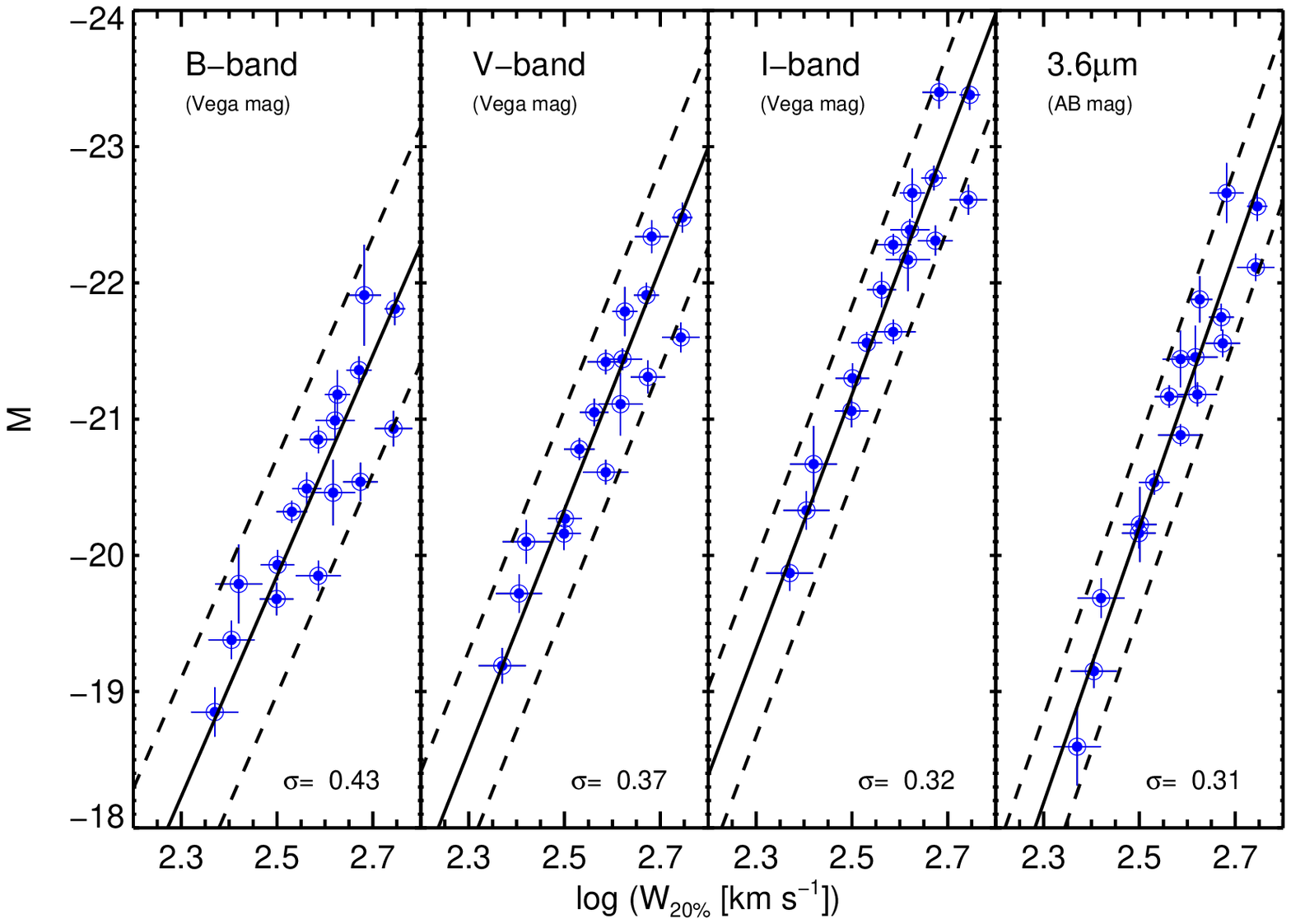}
\caption{\small Multi-wavelength Tully-Fisher relations. The three
  left panels show the B, V and I-band TF relations for galaxies
  calibrated with independently measured Cepheid moduli at the end of
  the HST Key Project. W is the inclination-corrected line width
  (from Sakai {\it et al.} 2000) measured at the 20\% power
  point. The right-hand panel shows the TF relation for the subset of
  galaxies drawn from the Key Project calibrators with measured 3.6
  $\mu$m total AB magnitudes from Seibert {\it et al.}  (2012). Data
  for 17 galaxies are available at all four wavelengths. The
  dispersions in these relations are shown in the lower right.}
\end{figure*}

\begin{figure*}
    \centering
    \includegraphics[width=10cm, angle=0]{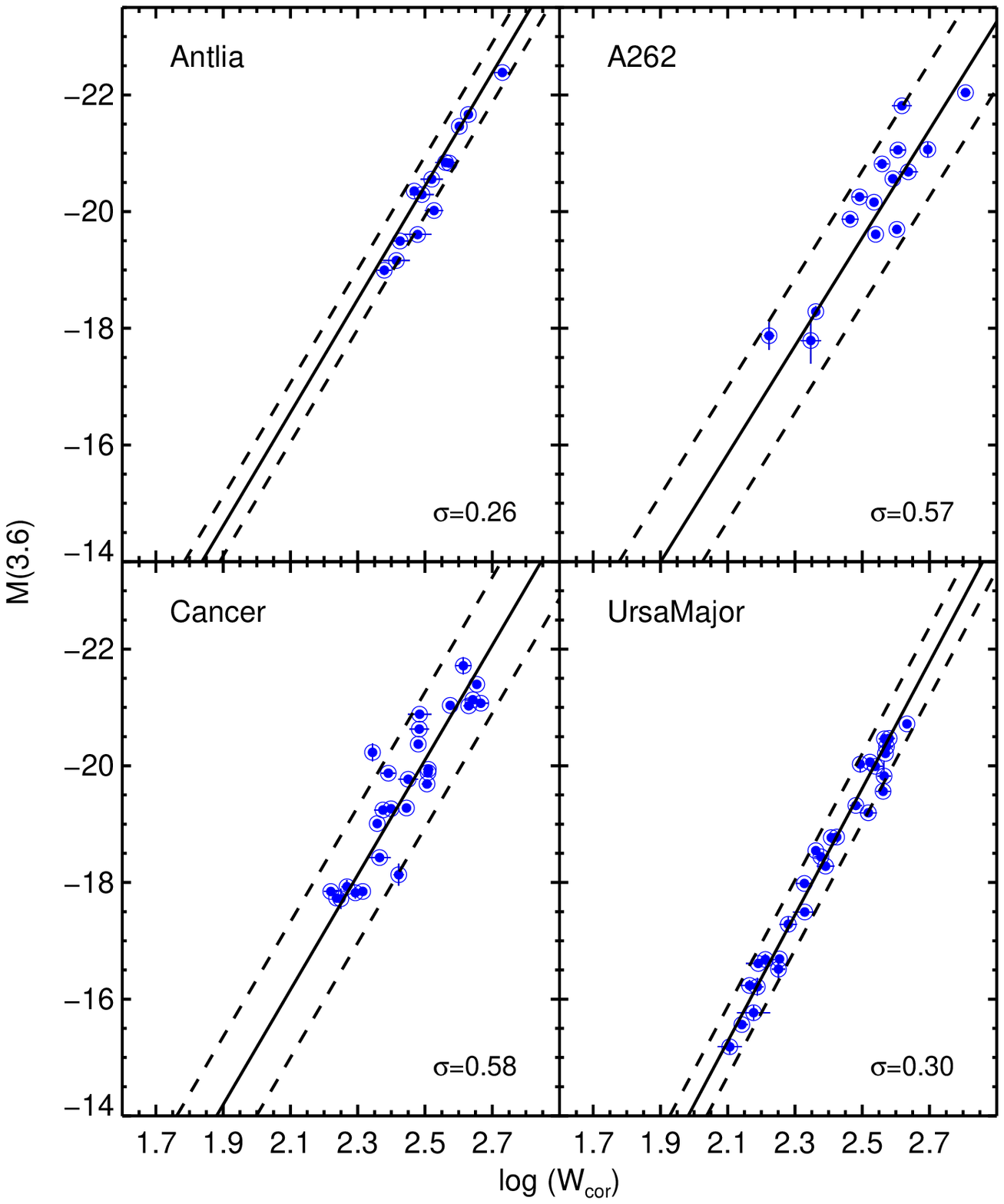}
\caption{\small Example 3.6 $\mu$m Tully-Fisher relations for four clusters of
  galaxies (Antlia, A262, Cancer and Ursa Major) from Seibert 
{\it et al.} (2012).  The dispersions in these relations are shown in the
  lower right.}
\end{figure*}

\section{Tests for Metallicity Sensitivity of the Mid-IR PL Relations}
\label{sec:metallicity}

A remaining systematic effect in the determination of $H_{\circ}$ is the
sensitivity of the Cepheid PL relation to metallicity.  We are
undertaking three independent tests of the sensitivity of the Cepheid
Leavitt Law  to metallicity; any one of which could, in
principle, calibrate the effect if it is measurable in the mid-IR,
and all three of which combined, will robustly constrain the
effect. The significant advantage of the mid-IR is that the relative
insensitivity to extinction allows a more precise test of metallicity
alone.

The first test involves the LMC alone (Freedman et
al. 2011). Romaniello{\it ~et al.}(2008) present evidence that the LMC
Cepheids themselves have a spread of metallicity amounting to 0.5~dex
in [Fe/H]. The mid-IR PL relations are predicted to have a residual
dispersion of less than $\pm$0.08~mag after time-averaged magnitudes
are obtained and geometrical effects due to the three-dimensional
extent and orientation of the LMC are removed. Any metallicity effect
must be buried within that small dispersion, along with any
other second and third-order effects, i.e., variations of radius and
temperature across the instability strip at fixed period, the presence
or absence of (physical) red companions, plane-thickness variations in
the LMC (over and above global tilt corrections) and residual
differential extinction effects.

In Figure 7, we show the deviations of individual JHK, 3.6 and 4.5
$\mu$m LMC Cepheid magnitudes from the period-luminosity relation as a
function of spectroscopic [Fe/H] metal abundances from Romaniello et
al. (2008).  [Fe/H] values range from approximately -0.6 to -0.1
dex. This plot is an updated version of Figure 2 from Freedman \&
Madore (2011), now based on time-averaged magnitudes, rather than the
two (random-phase) observations available previously. The slopes are
very shallow for all of the near- and mid-IR wavelengths; a crossover
occurs at the K-band where the slope is nearly flat.  Formally the 3.6
$\mu$m slope is -0.09 $\pm$ 0.29 mag/dex. We have also added data to this
plot for three Galactic Cepheids for which there are both 3.6 $\mu$m
data from Monson {\it et al.} (2012) and [Fe/H] measurements from
Romaniello {\it et al}. The Galactic Cepheid $l$ Car has a higher
[Fe/H] abundance than any of the LMC Cepheids with [Fe/H] = 0.0. Yet
there is again no indication in this (small) sample for a significant
metallicity effect. The slope is close to flat also for the Galaxy.

\begin{figure*}
\centering 
\includegraphics[width=10.0cm,  angle=-90]{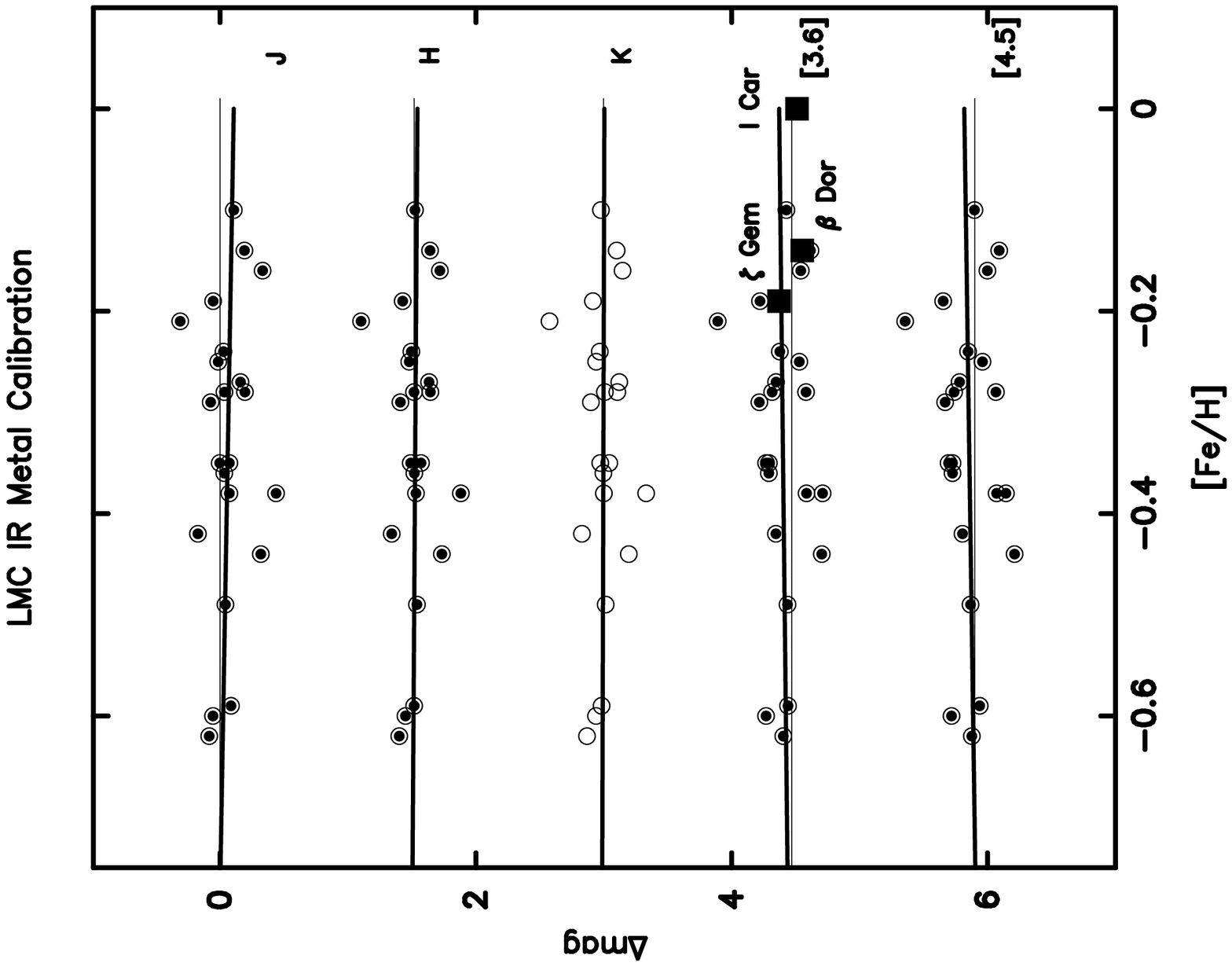}
\caption{ Deviations of individual JHK, 3.6 and 4.5 $\mu$m LMC Cepheid
  magnitudes from their respective period-luminosity relations plotted
  as a function of individual spectroscopic [Fe/H] metal abundances
  from Romaniello et al. (2008).  This is an updated and augmented
  version of a plot that first appeared in Freedman \& Madore (2011);
  here we only focus on the near-infrared (JHK) and mid-IR (3.6 and
  4.5 $\mu$m) relations.  The mid-IR LMC data are now based on
  time-averaged IRAC data (Scowcroft 2012a). Fits to the data are
  shown as thick solid lines; the thin solid lines are flat. For the
  3.6 $\mu$m correlation only we show the corresponding deviations
  (labeled solid squares) for three Galactic (high-metallicity)
  Cepheids which are common to the CHP and Romaniello samples. These
  data support the conclusion that metallicity effects at near- and
  mid-IR wavelengths are small compared to the optical and ultraviolet.  }
\end{figure*}

The second test involves both M31 and M33. Each galaxy supports a
modest metallicity gradient as measured from spectroscopic studies of
their HII regions. Given the insensitivity to reddening in the mid-IR,
a change in the apparent zero-point of the mid-IR PL relations as a
function of radius will provide a much stronger test for metallicity
than can be achieved at optical wavelengths alone where the combined
effects of reddening and metallicity are difficult to disentangle. We
have observed Cepheids in M31 and M33 covering the full range of
metallicity that each of these galaxies presents, sampled in at least
four radially distinct positions.

Finally, all targeted Local Group galaxies have independently
determined tip of the red giant branch (TRGB) distances.  By comparing
the TRGB distances with the mid-IR Cepheid distances one can test for
a correlation of those differences with metallicity (e.g., Lee,
Freedman \& Madore 1993; Sakai{\it ~et al.} 2004).  Cepheids and TRGB stars are
decoupled in their systematics, metallicities, history of star
formation and location in each of the galaxies.  They are of equally
high precision as distance indicators, and the moduli independently
derived from them can be compared.  We have observed Cepheids in
Sextans A and Sextans B, NGC~3109 and WLM, IC~1613 and NGC~6822 for
this test, obtaining twelve phase points per object in order
to bring the uncertainty in their individual mean mid-IR magnitudes
down to better than $\pm$0.04~mag.

We again emphasize that the strength of these tests resides in the
fact that the mid-IR Cepheid moduli require extremely small
corrections for extinction, unlike the same tests performed at optical
wavelengths where extinction corrections are far larger than the
sought-after metallicity effect.  If a metallicity effect exists at
mid-infrared wavelengths, we can calibrate and correct for it.

\section{Cepheid Mid-IR Colors: CO Absorption at  4.5 $\mu$m}
\label{sec:CO}

As noted above, the Rayleigh Jeans portion of the spectral energy
distribution for Cepheids is emitted at mid-IR wavelengths where there
is very little or no sensitivity to temperature. The slope of the tail
of the distribution is constant, independent of the temperature of the
star. Thus, if a Cepheid approximates a blackbody, the expectation
would be that colors based on the IRAC filters (at 3.5, 4.5, 5.8 or
8.0 $\mu$m) would be relatively constant as a function of phase and/or
period.  However, as seen previously in Figure 1, for the 4.5 $\mu$m
band, the presence of broad CO molecular absorption bands between
about 4 and 6 $\mu$m affects the 4.5 (and 5.8) $\mu$m IRAC filters
The 3.6 $\mu$m band lies outside of the CO
feature.  The models of Marengo {\it et al.}  (2010) show that these
deep CO absorption bands occur in all supergiants of the temperature
and gravity of Cepheids, and they also vary during the Cepheid
pulsational cycle. The color variation seen through the cycle results
from the fact that the CO absorption is sensitive to temperature and
varies both within a single Cepheid's pulsation cycle and between
Cepheids of different mean temperatures.

In Figure ~2, we showed 3.6 and 4.5 $\mu$m light curves, in addition
to [3.6]-[4.5] $\mu$m color curves for two Cepheids in each of the
Milky Way, LMC and SMC.  In general, we find that the [3.6]-[4.5]
$\mu$m mid-IR color curves for most of the longer-period (P $>$ 10
day) Cepheids in the LMC and the Galaxy display a significant cyclical
variability.  To our knowledge, this effect has never been observed
previously since light curves for Cepheids at 4.5 $\mu$m have never
been obtained before (previous observations have been one or two
epochs only).  However, V Cen and HV12452, with P=5.5 and 8.7 days,
respectively, show little variability and have colors of zero. The
light curves for the SMC, which has a lower metallicity than the
Galaxy and the LMC, show very little effect at any period. As we have
discussed, this variability occurs as a result of the presence of the
CO bandhead falling within the 4.5 $\mu$m filter. The CO feature
strengthens when the stellar atmosphere is more expanded and therefore
cooler. As seen in Scowcroft {\it et al.} the amplitude of the color
variability also increases with increasing period. No cyclical CO
variability is seen for Cepheids with periods less than about 10 days
(the hottest Cepheids). A detailed discussion of the 4.5 $\mu$m CO
feature in our Cepheid sample is presented in Scowcroft et
al. (2012c).

\section{Reducing the Uncertainty in H$_\circ$}
\label{sec:H0}

At the end of the Key Project, the overall
systematic uncertainty in the value of the Hubble constant was found
to be 10\% (Freedman et al. 2001).  Three of the largest sources of
systematic uncertainty listed included (a) 5\% involving the distance
to the Large Magellanic Cloud, setting the zero point of the Cepheid
PL relation, (b) 3.5\% due to the uncertainties involved in making the
photometric tie-in between ground-based telescopes and the HST
photometric system(s) and (c) $\pm$4\% uncertainty due to the difference in
metallicity between the LMC and the higher-metallicity spiral galaxies
in the Key Project sample.

All three of these major systematics are directly addressed by the use
of {\it Spitzer}. The mid-infrared data, which now include Galactic
zero-point calibrators, and rich sampling of long-period LMC Cepheids
defining the slope and width of the mid-IR PL relation, have been made
through the identical 3.6 $\mu$m filter, using the same instrument
(IRAC), on the same, stable platform (``Warm {\it Spitzer}''). In doing so, the
photometric tie-in uncertainties have, by design, been
eliminated. Similarly, the shape, width and zero-point calibration of
the Cepheid PL relation are also now well-measured using a combination
of HST parallaxes and {\it Spitzer} observations.  Moreover, the Milky Way
Galaxy has a metallicity comparable to those of the bulk of the HST
Key Project sample of spiral galaxies, thereby reducing  the
uncertainty previously incurred in using the (lower-metallicity) LMC as the zero-point
calibration.

Figure ~8 shows the 3.6 $\mu$m Leavitt Law for the five Galactic
Cepheids from Monson {\it et al.} (2012) with log P $>$ 0.8 and with
measured HST parallaxes from Benedict {\it et al.} (2007).  These are
$l$ Car, $\zeta$ Gem, $\beta$ Dor, W Sgr and X Sgr. Shown also are 82 LMC
Cepheids from Scowcroft {\it et al.} (2012a), with 0.8 $<$ log P $<$
1.8.  A new calibration of H$_\circ$ based on these data is presented
in (Freedman {\it et al.} 2012). The data are plotted to minimize the
variance in the combined sample. Error bars are smaller than the
plotted symbols. To maximize the overlap in the period range of the
Milky Way and LMC Cepheid samples, and to avoid uncertainties caused
by a possible difference in the slope of the Leavitt Law at short
periods, as well as to avoid overtone pulsators, only the five Milky
Way Cepheids with log P $>$ 0.8 are included in the fit.  The Galactic
calibrators (large symbols) fit within the scatter of the LMC Cepheid
Leavitt relation (small symbols), with a slope that is consistent, to
within the uncertainties, with that for the LMC.  The Galactic
Cepheids set the zero point of the relation; the larger sample of LMC
Cepheids is used to set the slope.

\begin{figure*}
    \centering
  \includegraphics[width=14.0cm, angle=-90]{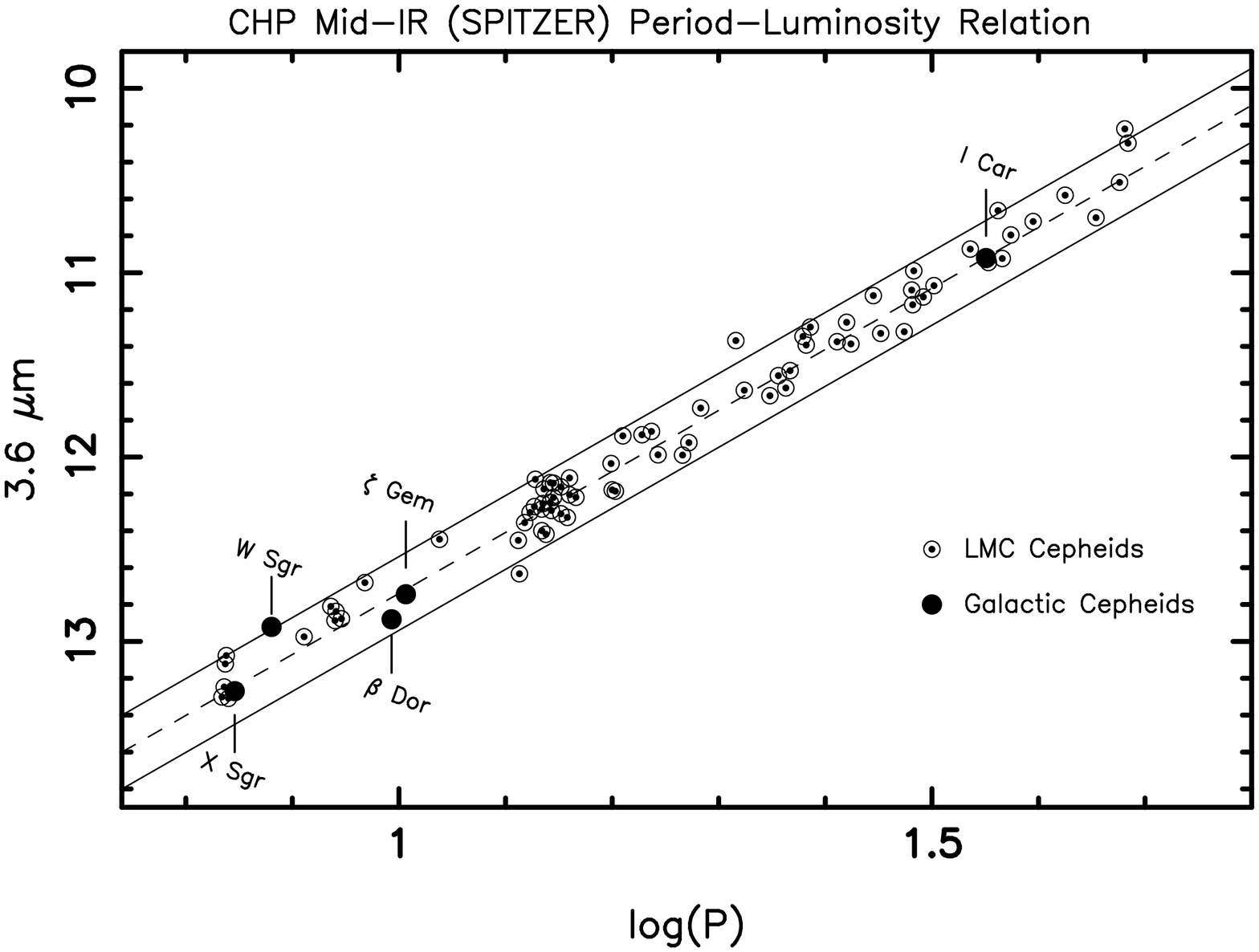}
\caption{Absolute calibration of the mid-infrared PL relation at 3.6 $\mu$m for
82 LMC Cepheids by five Milky Way Cepheids with trigonometric
parallaxes. Small circled points are LMC Cepheids; large filled circles
are Galactic Cepheids. }
\end{figure*}

\section{Summary}

A fundamental recalibration of the extragalactic distance scale is
underway using 3.6 $\mu$m observations of Cepheid variables to
determine distances within the Milky Way, throughout the Local Group,
and into the nearby and then more distant Hubble flow. This program is
explicitly designed to address directly all of the known local
systematic uncertainties currently impacting the optically-based
Cepheid distance scale.  As discussed in detail in Freedman et
al. (2012) these new data already result in a decrease of the
systematic uncertainty to $\pm$3\%, a factor of three over the Hubble
Key Project. Given the history of large systematic errors plaguing
attempts to measure an accurate value of the Hubble constant, this
small uncertainty may seem optimistic. However, this formal
uncertainty reflects the fact that the largest systematic
uncertainties (absolute zero point, the metallicity difference of the
LMC compared with the more distant spiral galaxies containing
Cepheids, reddening) have all been mitigated with {\it Spitzer} mid-IR
photometry. 


The major recent breakthroughs are the IRAC and {\it Spitzer}
observations of Cepheids at 3.6 $\mu$m, combined with the
determinations of direct high-precision geometric parallaxes to
Galactic Cepheids (Benedict {\it et al.} (2007). These data provide a
solid zero point (from the parallaxes) and a very high-precision
distance indicator (using the mid-infrared Leavitt Law slope and
scatter defined by LMC Cepheids). The mid-IR data are almost
completely insensitive to each of the previously known systematic
uncertainties that ultimately limited the previous determination of
H$_o$ using Cepheids to a 10\% uncertainty.

The limiting factor then becomes the small number of Galactic Cepheids
that have absolute parallaxes sampling the finite width of the PL
relation. We have therefore identified the next tier of Cepheids that
will be parallax targets for GAIA. As part of  the CHP, the mid-IR data 
are now already in hand for these calibrators, and awaiting parallax measurements.  At that
point the systematic error on the Galactic zero point will have been retired  to
$\pm$1\% allowing the total uncertainty on the Hubble constant to drop to
$<\pm$2\%.

This work is based in part on observations made with the {\it Spitzer}
Space Telescope, which is operated by the Jet Propulsion Laboratory,
California Institute of Technology under a contract with NASA. Support
for this work was provided by NASA through an award issued by
JPL/Caltech. We thank the staff of the {\it Spitzer} Science Center
for their help with the analysis of these data.

\vfill\eject

\vfill\eject

\clearpage 

\begin{table}
\caption{CHP {\it Spitzer} observations} 
\label{tab2}
\begin{tabular}{lccc}
\\ \hline
Program & Target & N$_{obs}$ & Channels \\ \hline
Milky Way & 37 Cepheids & 24 phased per Cepheid & [3.6], [4.5]  \\
LMC & 85 Cepheids & 24 phased per Cepheid & [3.6], [4.5]  \\
SMC & 100 Cepheids & 12 phased per Cepheid & [3.6], [4.5]  \\ \hline
Other Local Group & IC 10 & 6 random & [3.6], [4.5] \\
 & IC 1613 & 12 random & [3.6], [4.5] \\
 &  Leo A & 6 random & [3.6], [4.5] \\
  & M31, two fields & 12 random & [3.6], [4.5] \\
 & M33 & 12 random & [3.6], [4.5] \\
 & NGC 3109 & 12 random & [3.6], [4.5] \\
 & NGC 6822 & 12 random & [3.6], [4.5] \\
& Sextans A & 12 random & [3.6], [4.5] \\
& Sextans B & 12 random & [3.6], [4.5] \\
 & Pegasus Dwarf & 12 random & [3.6], [4.5] \\
 & WLM & 12 random & [3.6], [4.5] \\ \hline
Beyond the Local Group  & GR8 &  6 random & [3.6], [4.5] \\
& IC 4182 & 6 random & [3.6], [4.5] \\
& NGC 5253 & 6 random & [3.6], [4.5] \\
& M81, two fields & 8 random & [3.6] \\
& NGC 247 & 10 random & [3.6] \\
& NGC 300, 2 fields & 5 random & [3.6] \\
& NGC 7793 & 9 random & [3.6] \\
& NGC 2403, 2 fields & 5 random & [3.6] \\
& M101 & 5 random & [3.6] \\
& NGC 4258 & 12 random & [3.6] \\
& Cen A & 8 random & [3.6] \\
& M83 & 8 random & [3.6] \\ \hline
Tully--Fisher calibrators & 5 galaxies & 1 per galaxy & [3.6], [4.5] \\
Tully--Fisher targets & 398 targets & 1 per target &  [3.6], [4.5] \\
Supernova host galaxies & 44 targets & 1 per target & [3.6], [4.5] \\ \hline 
\end{tabular}
\tablecomments{Warm {\it Spitzer} observations taken in programs P61000 - 61010, and P70010 (SMC) for the Carnegie Hubble Program.}
\end{table}

\clearpage

\vfill\eject
\section{References}
\medskip

\par\noindent
Alves, D. R., 2004, New Astron. Rev., 48,  659


\noindent
Benedict, F. ~et al., 2007, AJ, 133, 1810

\noindent
Beulter, F., Blake, C., Colless, M., et al. 2011, MNRAS, in press (arXiv:1106.3366)

\noindent
Bonanos, A., \& Stanek, C., 2003, ApJ, 591,  L111 

\noindent
Braatz, J. A. \& Gugliucci, N. E. 2008. ApJ, 678, 96

\noindent
Contreras, C., ~et al.\ 2010, AJ, 139, 519 

\noindent
Dale, D.A., ~et al., 2007, ApJ, 655,  863 

\noindent
Fazio, G.G., ~et al., 2004, ApJS, 154, 10

\noindent
Ferraresse, L., {\it ~et al.} 2007, ApJ., 654, 186

\noindent
Flaherty, K. M.{\it ~et al.} 2007, ApJ, 663, 1069

\noindent
Freedman, W. L., Rigby, J., Madore, B.F., Persson, S.E., Sturch, L. V. \&
Mager, V., 2009, ApJ, 695, 996

\noindent
Freedman, W. L., ~et al., 2001, ApJ, 553, 47

\noindent
Freedman, W. L., \& Madore, B. F., 2010, ARAA, 48, 673

\noindent
Freedman, W. L., Madore, B. F., Rigby, J., Persson, S.E., \& Sturch, L.  2008, ApJ, 679, 71

\noindent
Freedman, W.L., \& Madore, B.F. 2011, ApJ, 734, 46

\noindent
Freedman, W.L., ~et al., 2012, ApJL, in preparation


\noindent
Gibson, B. K. 2000, Mem. Soc. Astron. Italiana, 71, 693

\noindent
Gieren, W.,{\it ~et al.} 2008, ApJ., 683, 611

\noindent
Giovanelli, R., Haynes, M.~P., Herter, T., Vogt, N.~P., Wegner, G., Salzer, J.~J., da 
Costa, L.~N., \& Freudling, W., 1997, AJ, 113, 22 

\noindent
Hamuy, M., ~et al., 2006, PASP, 118,  2 

\noindent
Hu, W., 2005,  ASP Conf.Ser., 399, 215

\noindent
Hu, W., \& Dodelson, S. 2002, ARAA,  40,  171

\noindent
Humphreys, E.~M.~L., Reid, M.~J., Greenhill, L.~J., Moran, J.~M., 
\& Argon, A.~L.\ 2008  ApJ, 672, 800 

\noindent
Indebetouw, R.{\it ~et al.} 2005, ApJ, 619, 931

\noindent
Karachentsev, I.D., Karachentseva, V.E., \& Parnovsky, S.L., 1993, A.N., 314, 97

\noindent
Komatsu, E., Smith, K.M., Dunkley, J., et al. 2011, ApJS, 192, 18

\noindent
Kurucz, R. 1993, ATLAS9 Stellar Atmosphere Programs and 2 km/s
grid. Kurucz CD-ROM No. 13. Cambridge, Mass.: Smithsonian Astrophysical
Observatory, 1993., 13

\noindent
Leavitt, H. 1908, Ann. Harv. Coll. Obs. 60, 87

\noindent 
Lee, M.-G., Freedman, W. L., \& Madore, B. F.,  1993, ApJ, 417, 553

\noindent
Macri, L.~M., Stanek,  K.~Z., Bersier, D., Greenhill, L.~J., \& Reid, M.~J.  2006, ApJ, 652, 1133 

\noindent
Madore, B. F., \& Freedman, W., 1991,  PASP,  103, 933

\noindent
Madore, B. F., Freedman, W.L., Rigby, J., Persson, S.E., Sturch, L. V., \& Mager, V. 2009, ApJ, 695, 988

\noindent
Mager, V.~A., Madore, B.~F., \& Freedman, W.~L.  2008, ApJ, 689, 721 

\noindent
Marengo, M., Evans, N.~R., Barmby, P., Bono, G., Welch, D.~L., 
\& Romaniello, M.  2010, ApJ, 709, 120 

\noindent
Meixner, M.,  ~et al., 2006, AJ, 132, 2268

\noindent 
Monson, A.J., Scowcroft, V., Rigby, J., Freedman, W.L., Madore, B.F.,
Persson, S.E., Seibert, M., Stetson, P., \& Sturch, L. V., 2012, ApJ, submitted

\noindent 
Mould, J.R. 2011, PASP, in press (arXiv:1107.2982)

\noindent
Nishiyama, S., {\it ~et al.} 2009, ApJ, 696, 1407

\noindent
Persson, S.E., ~et al., 2004, AJ, 128, 2239

\noindent
Rieke, G.~H., \& Lebofsky, M.~J. 1985, ApJ, 288, 618

\noindent
Riess, A.~G.,  ~et al., 2005, ApJ, 627, 579

\noindent
Riess, A.~G.,  ~et al., 2009, ApJ, 699, 539

\noindent
 Riess, A.~G., et al.\  2011, ApJ, 730, 119 

\noindent
Roman-Zuniga, C.~G., {\it ~et al.} 2007, ApJ, 664, 357

\noindent
Romaniello,  M.,  et al.   2008, A\&A, 488, 731

\noindent
Sakai, S., ~et al., 2000, ApJ, 529, 698

\noindent
Sakai, S.,{\it ~et al.} 2004, ApJ, 608, 42

\noindent
Sandage, A.R., Tammann, G.A., Saha, A., Reindl, B., Macchetto,
F.~D., \& Panagia, N. 2006, ApJ, 653, 843

\noindent
Sbordone, L., Bonifacio, P., Castelli, F., \& Kurucz, R. L. 2004, Memorie
della Societa Astronomica Italiana Supplement, 5, 93

\noindent
Sbordone, L., 2005, Memorie della Societa Astronomica Italiana Supplement, 8, 61

\noindent
Schaefer, B.E., 2008, AJ, 135, 112

\noindent 
Scowcroft, V.,  Freedman, W.L., Madore B.F., Monson, A.J.,
Persson, S.E., Rigby, J., Seibert, M., Stetson, P., \& Sturch, L. 2012a, ApJ, submitted

\noindent 
Scowcroft, V.,{\it ~et al.} 2012b, ApJ, in preparation

\noindent 
Scowcroft, V.,{\it ~et al.} 2012c, ApJ, in preparation

\noindent
Seibert, M., ~et al., 2012, in preparation

\noindent
Spergel, D.N.,{\it ~et al.} 2007, ApJ, 170, 377

\noindent
Turner, D.G. 2010, Ap \& SS , 326, 219

\noindent
Werner, M. W.,{\it ~et al.} 2004, ApJS, 154, 1

\end{document}